
\magnification=1200

\documentstyle{amsppt}

\language=0			



\def\SLZ{\operatorname{SL}(2,\Z)}
\def\SL{\Gamma}			\def\GL{\operatorname{GL}}


		\def\Z{{\Bbb Z}}
\def\Q{{\Bbb Q}}		\def\C{{\Bbb C}}
\def\F{{\Bbb F}}		\def\FF{\F{(l^2)}}


\def\O{{\frak O}}		\def\o{{\frak o}}
\def\p{{\frak p}} 		


	\def\Im{\operatorname{Im}}
\def\norm{\operatorname{n}}	\def\tr{\operatorname{tr}}
\def\id{\operatorname{id}}	\def\e{\operatorname{e}}
\def\QF{{\Cal Q}}		\def\BF{{\Cal B}}
\def\diag{\operatorname{diag}}


\def\PH{\frak H}		\def\W{{\Cal W}}
\def\BO{{\Cal O}}


\def\BEHHH{B-E-H$^3$}
\def\ES{E-S}
\def\FreKaWa{F-K-W}
\def\GP{GP}
\def\HuseMil{H-M}
\def\Kac{K}
\def\Nobs{N}
\def\NW{N-W}
\def\RoCa{R-C}
\def\Schoen{Sch}
\def\Vi{V}


\font\HUGE=cmbx12 scaled \magstep4
\font\Huge=cmbx10 scaled \magstep4

\font\large=cmr17 scaled \magstep0

\document
\nopagenumbers
\pageno = 0
\centerline{\HUGE Universit\"at Bonn}
\vskip 10pt
\centerline{\Huge Physikalisches Institut}
\vskip 2.5cm
\vskip 1.4cm
\centerline{\large Wolfgang Eholzer\raise7pt\hbox{\rm 1} and
                   Nils-Peter Skoruppa\raise7pt\hbox{\rm 2} }
\vskip 1.8cm
\centerline{\bf Abstract}
\vskip3true mm
\noindent
We describe the construction of vector valued modular forms
transforming under a given congruence representation of the
modular group $\SLZ$ in terms of theta series.
We apply this general setup to obtain closed and easily computable
formulas for conformal characters of rational models of
$\W$-algebras.
\vskip 1.8cm
\line{\hbox to 5cm{\hrulefill} \hfill}
\line{${}^{2}$
Universit\'e Bordeaux I,
UFR de Math\'ematiques et Informatiques,
\hfill}
\line{\phantom{${}^{2}$}
351 rue de la Lib\'eration,
33405 Talence, France
\hfill}
\line{\phantom{${}^{2}$} e-mail: skoruppa\@ceremab.u-bordeaux.fr\hfill}
\vskip 1.5cm
\settabs \+&  \hskip 110mm & \phantom{XXXXXXXXXXX} & \cr
\+ & ${}^{1}$ Post address:                & BONN--TH--94--24& \cr
\+ & Nu{\ss}allee 12                       & MSRI            & \cr
\+ & D-53115 Bonn                          & hep-th/9410077  & \cr
\+ & Germany                               & Bonn University & \cr
\+ & e-mail:                               & Oktober 1994    & \cr
\+ & eholzer\@mpim-bonn.mpg.de             &                 & \cr
\vfill
\eject


\topmatter
	\address
Max-Planck-Institut f\"ur Mathematik Bonn,
Gottfried-Claren-Stra{\ss}e 26,
53225 Bonn, Germany
	\endaddress
\email eholzer\@mpim-bonn.mpg.de \endemail
	\address
Physikalisches Institut der Universit\"at Bonn,
Nu{\ss}allee 12,
53115 Bonn, Germany
	\endaddress
\email eholzer\@avzw01.physik.uni-bonn.de \endemail
	\address
Universit\'e Bordeaux I,
U.F.R\. de Math\'ematiques et Informatique,
351 rue de la Lib\'eration,
33405 Talence, France
	\endaddress
\email skoruppa\@ceremab.u-bordeaux.fr \endemail
	\abstract
We describe the construction of vector valued modular forms
transforming under a given congruence representation of the
modular group $\SLZ$ in terms of theta series.
We apply this general setup to obtain closed and easily computable
formulas for conformal characters of rational models of
$\W$-algebras.
	\endabstract
	\toc
\head
1. Introduction
\endhead
\head
2. An axiomatic characterization of the conformal characters of five
   special rational models
\endhead
\head
3. Realization of modular representations by theta series
\endhead
\head
4. Theta series associated to quaternion algebras and the conformal
   characters of the five special models
\endhead
\head
5. Comparison to formulas derivable from the representation theory of
   Kac-Moody and Casimir $\W$-algebras
\endhead
	\endtoc
\endtopmatter

\rightheadtext{	Conformal Characters and Theta Series}
\leftheadtext{W\. Eholzer and N-P\. Skoruppa}

\document

\head
1. Introduction
\endhead
Once a rational model of a $\W$-algebra is constructed or its existence
is conjectured the next natural step is to compute explicitly its conformal
characters. This is interesting for giving further evidence to the
existence of the rational model in question, if its existence
is not sure, and it is interesting in that conformal characters are usually
surprisingly distinguished modular functions which are often related
to a priori purely number theoretical objects like generalized
Rogers-Ramanujan identities or dilogarithm identities.

However, in general it is quite difficult to construct conformal characters
directly from physical information about the $\W$-algebra; even the direct
computation of their first few Fourier coefficients affords considerable
computer power.

In \cite{\ES} we took a different point of view in order to determine
conformal characters amongst all possible modular functions.

We formulated a list of certain axioms which hold true for all known
sets of conformal characters of rational models of $\W$-algebras. The only
data from an underlying rational model which occur in these axioms are
respectively its central charge and its conformal dimensions.
We showed that, for several rational models, these axioms uniquely
determine the conformal characters belonging to a given list of central charge
and conformal dimensions.

Thus, once the central charge and conformal dimensions of a rational model
are known, the computation of its conformal characters can be viewed as
a problem which is completely independent from the theory
of $\W$-algebras, i.e\. for this computation one is left with a construction
problem, namely, the problem
of finding, by whatever means, a set of functions fulfilling the indicated
list of axioms.

The purpose of the present note is to describe such a mean which can solve in
many
cases this construction problem. In particular, we shall apply our method to
the case of five special rational models. The reason for the choice of these
models is that the representation theory of the $\SLZ$-representation on their
conformal characters can be treated homogeneously in some generality, and that
the conformal characters of one of these models
(of type $\W(2,8)$ with central charge $c=-\frac{3164}{23}$)
could not be computed by the so far known methods.

This note is organized as follows:
In section 2 we give a short description of the five special models.
We recall the axiomatic description of conformal characters, and, in
particular, we
recall the results obtained in \cite{\ES} which concern the five special models
{}.
In Section 3 we describe a general procedure for
the construction of vector valued modular forms transforming under
a given matrix representation of $\SLZ$. As already mentioned,
this procedure is thought to be useful in general
for finding explicit and easily computable formulas for conformal characters.
In section~4 we
apply the general setup of the preceding section to the case of the five
special models,
and we derive explicit formulas for their conformal characters. Finally, in \S
5 we compare
our results with those formulas for the conformal characters of the five models
which can be obtained (assuming  certain conjectures) from the representation
theory
of Casimir-$\W$-algebras.

\subhead
Notation
\endsubhead
We use $\PH$ for the complex upper half plane,
$\tau$ as a variable in $\PH$,
$q= e^{2\pi i \tau}$,
$q^\delta=e^{2\pi i \delta\tau}$,
$T=\left(\smallmatrix 1&1\\0&1\endsmallmatrix\right)$,
$S=\left(\smallmatrix 0&-1\\1&0\endsmallmatrix\right)$,
$\SL$ for the group $\SLZ$, and
$$\Gamma(n) = \{ A \in \SLZ \ \vert \ A \equiv \id \pmod n \}$$
for the principal congruence subgroup  of $\SLZ$ of level $n$.
We use  $\eta$ for the Dedekind eta function
$$\eta(\tau)=\e^{\pi i\tau/12}\prod_{n\ge 1}(1-q^n).$$
The group $\SL$ acts on $\PH$ by
$$A\tau=\frac{a\tau+b}{c\tau+d}\qquad
(A=\pmatrix a&b\\c& d\endpmatrix).$$
For a complex vector valued function $F(\tau)$ on $\PH$, and for an integer $k$
 we use
$F|_kA$ for the function defined by
$$(F|_kA)(\tau)=(c\tau+d)^{-k}F(A\tau).$$
Finally, for a matrix representation $\rho\colon\SL@>>>\GL(n,\C)$ and an
integer $k$ we use
$M_k(\rho)$ for the vector space of all holomorphic
maps $F\colon\PH\to\C^n$ (= column vectors) which satisfy $F|_kA=\rho(A)F$ for
all
$A\in\SL$, and which are bounded in any region $\Im(\tau)\ge r>0$.
Thus, if $\rho$ is the trivial representation, then $M_k(\rho)$ is the space of
ordinary
modular forms on $\Gamma$ and of weight $k$.

\head
2. An axiomatic characterization of the conformal characters of five special
rational models
\endhead

In this section we recall some results on the conformal characters of 5
rational models,
which were, among others, considered in \cite{\ES}. The types, the central
charges,
the sets of conformal dimensions $H_c$, and the effective central charge
$\tilde c=c-24\min H_c$ of these models are listed in Table 1;
for a more detailed description the reader is referred to loc\. cit..
Denote by $\chi_{c,h}$ ($h\in H_c$) the conformal characters
of the rational model with central charge $c$, where we choose the notation
such that
$$\chi_{c,h}=q^{h-c/24}\cdot(\text{power series in }q).$$

\medskip
\centerline{Table 1:  Central charges and conformal dimensions}
\smallskip
\centerline{\vbox{\offinterlineskip
\def\tablespace{height2pt&\omit&&\omit&&\omit&&\omit&\cr}
\def\tablerule{\tablespace\noalign{\hrule}\tablespace}
\hrule
\halign{&\vrule#&\strut\quad\hfil#\hfil\quad\cr
\tablespace
&type&&$c$&&$\tilde{c}$&&$H_c$&\cr
\tablerule
&$\W_{G_2}(2,1^{14})$&&$-{8\over5}$&&${16\over5}$&&${1\over5}\{0,-1,1,2\}$&\cr
\tablerule
&$\W_{F_4}(2,1^{26})$&&${4\over5}$&&${28\over5}$&&${1\over5}\{0,-1,2,3\}$&\cr
\tablerule
&$\W(2,4)$&&$-{444\over11}$&&${12\over11}$&&$-{1\over11}\{0,9,10,12,14,15,16,17,18,19\}$&\cr
\tablerule
&$\W(2,6)$&&$-{1420\over17}$&&${20\over17}$&&$-{1\over17}\{0,27,30,37,39,46,48,49,50,$&\cr
&\omit&&\omit&&\omit&&\qquad$52,53,55,57,58,59,60\}$&\cr
\tablerule
&$\W(2,8)$&&$-{3164\over23}$&&${28\over23}$&&$-{1\over23}\{0,54,67,81,91,94,98,103,111,$&\cr
&\omit&&\omit&&\omit&&\quad$112,116,118,119,120,122,124,$&\cr
&\omit&&\omit&&\omit&&\quad$125,129,130,131,132,133\}$&\cr
\tablespace}
\hrule}}
\medskip

Fix now one of the  central charges $c$ of Table 1.  Suppose that we are given
functions $\xi_{c,h}$ $(h\in H_c$), defined on the upper half plane $\PH$,
which satisfy
the following list of properties:

\definition{Properties of conformal characters}
\roster
\item
The functions $\xi_{c,h}$ are nonzero modular functions for some
congruence subgroup of $\SL=\SLZ$ without poles in the upper half plane.
\item
The
space of functions spanned by the $\xi_{c,h}$ ($h\in\ H_c$)
is invariant under $\SL$ with respect to the action
$(A,\xi)\mapsto\xi(A\tau)$.
\item
For each $h\in H_c$ one has $\xi_{c,h}=\BO(q^{-\tilde c/24})$
as $\Im(\tau)$  tends to infinity, where $\tilde c = c -24 \min H_c$.
\item
For each $h\in H_c$ the function $q^{-{(h-\frac c{24})}}\xi_{c,h}$
is periodic with period 1.
\item
The Fourier coefficients of the $\xi_{c,h}$ are rational numbers.
\endroster
\enddefinition

It was shown in \cite{\ES} that the functions $\xi_{c,h}$ are uniquely
determined by
these five properties, up to multiplication by scalars.
Very likely the conformal characters $\chi_{c,h}$ satisfy the properties (1) to
(5) if
we view them as functions of $\tau$ by setting $q=\exp(2\pi i\tau)$ (for a
detailed
discussion of this conjecture cf\. loc\. cit.). Thus, assuming this conjecture
for a
moment, we conclude that, for any set of functions $\xi_{c,h}$ as above, we
have
$\chi_{c,h}=\text{const.}\cdot \xi_{c,h}$.

As already explained the main purpose of this paper is the explicit
construction of functions $\xi_{c,h}$
satisfying (1) to (5). To this end we have to recall more precisely what we
showed in \cite{\ES}.

Denote by $\xi_c$ the column vector whose entries are the $\xi_{c,h}$ ordered
in some way.
Let $l$ be the denominator of $c$, let $k$ be the integer associated to $c$ by
\
Table 2,
and let $\rho_l$ be the $\SL$-representation
defined by
$$(\eta^{2k}\xi)|_kA = \rho_l(A)\,\eta^{2k}\xi\qquad(A\in\SL).$$
As we shall see in a moment the a priori two different representations with
$l=\
5$ are in fact equivalent; so for notational convenience we denote them by the
\
same symbols. We showed in \cite{\ES; cf\. \S  4.4} that properties (1) to (5)
imply

\proclaim{ Theorem (Characterization of $\rho_l$)}
The representation $\rho_l$ is
irreducible,
its kernel contains $\Gamma(l)$, and it takes its values in
$\GL(l-1,\Q(\e^{2\pi i/l}))$.
\endproclaim

Consulting any table of irreducible representations of
$\SL/\Gamma(l)\approx\operatorname{SL}(2,\Z/l\Z)$
one verifies that the three properties listed in the Theorem uniquely determine
$\rho_l$ up to equivalence.
We shall give an explicit description of $\rho_l$ in \S\ 4.

{}From property (3) we have $\eta^{2k}\xi_c=\BO(q^\delta)$ for $q\to 0$, where
$\delta = -\tilde c + k/12$, and, in particular,  that $\eta^{2k}\xi_c$ is an
element of $M_k(\rho_l)$. The dimensions of these spaces can be computed using
the dimension formula in \cite{\ES}. The resulting dimensions and the values of
$\delta$ are listed in Table 2.

Let $M^{(\delta)}_k(\rho_l)$ be the subspace of all $F\in M_k(\rho_l)$
satisfying $F=\BO(q^\delta)$. In \cite{\ES} it was shown that this subspace
is one-dimensional, which, by obvious arguments, implies that $\xi_c$ is
unique up to multiplication by diagonal matrices (actually, it was shown
in loc\. cit\. that $M_h(\rho_l\otimes \theta^{2h-2k})$
is one-dimensional, where $\theta^2(A)=(\eta^2|_1A)/\eta^2$. However, this
latter
space is obviously isomorphic to $M^{(\delta)}_k(\rho_l)$ via multiplication by
$\eta^{2k-2h}$).

\medskip
\centerline{ Table 2: Certain data related to the five rational models}
\smallskip
\centerline{%
\vbox{\offinterlineskip
\def\tablespace{height2pt&\omit&&\omit&&\omit&&\omit&&\omit&\cr}
\def\tablerule{\tablespace\noalign{\hrule}\tablespace}
\hrule
\halign{&\vrule#&\strut$\quad\hfil#\hfil\quad$\cr
\tablespace
&\W\text{-algebra}&&c&&l&&k&&\delta&&\dim M_k(\rho_l)&\cr
\tablerule
&\W_{G_2}(2,1^{14})&&-{8\over5}&&5&&4&&\frac15&&1&\cr
\tablerule
&\W_{F_4}(2,1^{26})&&{4\over5}&&5&&10&&\frac35&&3&\cr
\tablerule
&\W(2,4)&&-{444\over11}&&11&&6&&\frac5{11}&&5&\cr
\tablerule
&\W(2,6)&&-{1420\over17}&&17&&2&&\frac2{17}&&2&\cr
\tablerule
&\W(2,8)&&-{3164\over23}&&23&&10&&\frac{18}{23}&&17&\cr
\tablespace
}
\hrule}}
\medskip

\head
3. Realization of modular representations by theta-series
\endhead

In this section we show how one can, under certain hypothesis, construct
systematically vector
valued modular in $M_k(\rho)$ for a given matrix representation
$\rho$ of $\Gamma$ and given weight $k$.

The first step is a realization of $\rho$ as subrepresentation of a Weil
representation. We explain this notion:
Let $M$ be a finite abelian group and
$$\QF\colon M@>>>\Q/\Z$$
a non-degenerate quadratic form.
By quadratic form we mean that $\QF$ is even (i.e\. $\QF(-x)=\QF(x)$) and the
map
$$(x,y)\mapsto \BF(x,y):=\QF(x+y)-\QF(x)-\QF(y)$$
is $\Z$-bilinear, and by non-degenerate we mean that for
each $x\not=0$ there exists a $y$ such that $\BF(x,y)\not=0$.
We shall call such a pair $(M,\QF)$ a quadratic module.
The Weil representation $\omega=\omega_{(M,\QF)}$ associated to $M$ and $\QF$
is a
projective right-representation of $\SL$ on the space $\C^M$ of complex-valued
functions $f$ defined on $M$. The operators corresponding under $\omega$ to the
generators $S$ and $T$ of $\SL$ are given by
$$f|\omega(T)(x)=\e(\QF(x))\,f(x),\quad
f|\omega(S)(x)=\gamma(1)^{-1}\,\sum_{y\in
M}\e(-\BF(x,y))\,f(y).$$ Here, for any integer $a$
we use
$$\gamma(a)=\sum_{x\in M}\e(a\QF(x)).$$
Moreover, here and in the following, for any $x\in \Q/\Z$, we shall write
$\e(x)$
for $\e^{2\pi i r}$, where $r$ is any representative of $x$.
The reader is referred to \cite{\Nobs} for more details on Weil representations
(To translate our terminology to the one in \cite{\Nobs} note that
$f|\omega_{(M,\QF)}(A)=w_{(M,-\QF)}(A^{-1})f$, where $w_{(M,-\QF)}$ is the Weil
(left-)representation associated to the quadratic module $(M,-\QF)$ as in
\cite{\Nobs}.)

Denote by $l$ the level of $\QF$, i.e. the smallest
positive integer such that
$l\QF(M)=~0$. Using that $\QF$ is
non-degenerate it is easy to show that $|\gamma(a)|=|M|^{1/2}$ for any integer
relatively prime to $l$. Moreover, $\gamma(a)$ is
obviously an element of $K=\Q(\e^{2\pi i/l})$. We identify $(\Z/l\Z)^\ast$ with
the Galois
group $G$ of $K$ in the usual way such that $a\in (\Z/l\Z)^\ast$ corresponds to
the
Galois substitution $\sigma_a$ which is given by $\e(1/l)^{\sigma_a}=\e(a/l)$.
For
$a\in G$ set $co(a)=\gamma(a)/\gamma(1)$, and denote by $\mu_8$ the group of
eights
roots of unity. Then it is easy to show that $a\mapsto co(a)$ defines a
co-cycle of $G$
with values in $\mu_8$, i.e\. $co(a)$ is a eights root of unity and one has
$$co(a)^{\sigma_b}co(b)=co(ab)\qquad(a,b\in(\Z/l\Z)^\ast).$$

To verify this note that
$\gamma(a)=\gamma(1)^{\sigma_a}$, from which the co-cycle relation is obvious.
Moreover, this identity (together with $|\gamma(a)|=|M|^{1/2}$) implies
$|co(a)^\sigma|=1$ for all $\sigma\in G$, i.e\. that $co(a)$ is a root of
unity, say
$co(a)=\e(r/l)$. Since $co(a)$ is obviously invariant under $G^2$ we find that
$r(b^2-1)$, for any $b$ relatively prime to $l$, is divisible by $l$, which is
only
possible if $l/(l,r)$ divides $8$, i.e\. if $co(a)\in\mu_8$.

The Weil representation $\omega$ is a true representation, i.e\.
not only a projective
one, if and only if the co-cycle $co$ is a character \cite{\Nobs, Satz 1}.
If this is
the case we call $\omega$ a proper Weil representation. From the discussion
above it is
clear that $co$ is a character if and only if $co$ takes values in $\{\pm1\}$.
Moreover, if $\omega$ is proper, then its kernel contains $\Gamma(l)$
\cite{\Nobs, Satz 2}.

One now has the following theorem.
\proclaim{Theorem \cite{\NW}}
Each irreducible right-representation of $\Gamma$ whose kernel contains a
principal congruence subgroup is isomorphic to
a subrepresentation of a suitable proper Weil representation.
\endproclaim

We call two quadratic modules $(M,\QF)$ and $(M',\QF')$ isomorphic if there
exist an isomorphism (of abelian groups) $\pi\colon M@>>>M'$ such that
$\QF'\circ\pi=\QF$, and we denote such an isomorphism by
$$\pi\colon (M,\QF)@>\approx>>(M',\QF').$$ It is easy to show that
isomorphic quadratic modules yield isomorphic Weil representations: an
isomorphism of
(projective or proper) $\SL$-representations is given by the map
$$\pi^\ast\colon\C^{M'}@>>>\C^M,\qquad
f\mapsto \pi^\ast f=f\circ\pi.$$

As the next step for constructing elements of spaces $M_k(\rho)$
we connect Weil representations and theta series by lifting quadratic modules
to lattices and quadratic forms on them.

More precisely, let $(M,\QF)$ be a quadratic module. Assume that $L$ is a
complete
lattice in some rational finite-dimensional vector
space $V$ and $Q$ a positive definite non-degenerate quadratic form on $V$
which takes on integral values on $L$, and such that there exists an
isomorphism
of quadratic
modules
$$\pi\colon(L^\sharp/L,\widetilde Q)@>\approx>>(M,\QF).$$
Here we use
$L^\sharp$ for the dual lattice of $L$ with respect to $Q$, i.e\.
$L^\sharp$ is the set of all $y\in V$ such that
$B(L,y)\subset\Z\}$ with $B(x,y)=Q(x+y)-Q(x)-Q(y),$
and we use $\widetilde Q$ for the induced quadratic form
$$\widetilde Q:L^\sharp/L@>>>\Q/\Z\,,\qquad
x+L\mapsto Q(x) +\Z.$$
We shall call such a pair $(L,Q)$ a lift of the quadratic module
$(M,\QF)$.

Let $p$ a homogeneous spherical polynomial on $V$ with respect to $Q$ of
degree $\nu$,
i.e\. if we choose a basis $b_j$ of $V$, then $p\left(\sum b_j\xi_j\right)$
becomes a complex homogeneous polynomial in the
variables $\xi_j$ of degree $\nu$ satisfying
$$\nabla G^{-1}\nabla'\,p\big(\sum_j b_j\xi_j\big) = 0,$$ where
$\nabla=(\frac{\partial}{\partial\xi_1},\dots)$ and $G=(B(b_j,b_k))_{j,k}$ is
the
Gram matrix of $B$.

Finally, for $f\in\C^M$, set
$$\theta_f=\sum_{x\in L^\sharp}(\pi^\ast f)(x)\,p(x)\,q^{Q(x)}.$$
Here we view $\pi^\ast f$ as function on
function on $L^\sharp$ which is periodic with period lattice $L$.

We assume that $V$ has even dimension $2r$. Then the Weil representation
$\omega=\omega_{(M,\QF)}$
is proper as follows from the discussion of the co-cycle $co(a)$ above and
Milgram's theorem
which implies that $co(a)$ is $\pm1$ (cf\. Appendix below).
One has

\proclaim{Theorem (Representation by theta series)}
The map  $\C^M\ni
f\mapsto \theta_f$ has the property
$\theta_f|_{r+\nu}A=\theta_{f|\omega(A)}$ for all $A\in\SL$,
i.e\. it defines a homomorphism of $\SL$-modules.
\endproclaim

This is, in various different formulations, a well-known theorem. For the
reader's
convenience we shall
sketch the proof in the appendix below.

Let now $\rho\colon\SL@>>>\GL(n,\C)$ be a congruence
matrix representation, and assume that we have determined a
quadratic module $(M,\QF)$
such that the associated Weil representation is proper and contains a
subrepresentation which is isomorphic to the (right-)representation
$\C^n\times\SL\ni(z,A)\mapsto z\rho(A)'$, where the prime denotes
transposition.
The existence of such a $(M,\QF)$ is guaranteed by the first theorem.
Thus, there exists a $\SL$-invariant subspace of $\C^M$ with basis $f_j$
such that
$$\Phi|\omega(A)=\rho(A)\Phi\qquad(A\in\SL),$$ where
$\Phi$ denotes the column vector build from the $f_j$.
Assume furthermore that there exists a lift $(L,Q)$ of $(M,\QF)$, i.e\.
an isomorphism
$$\pi\colon(L^\sharp/L,\widetilde Q)@>\approx>>(M,\QF)$$
with a lattice $L$ of even rank $2r$.
Let $p$ be a homogeneous spherical polynomial w.r.t\. $Q$ of degree $\nu$.
{}From the last theorem it is then clear that we have the
following

\proclaim{Theorem (Realization by theta series)} The function
$$\theta=\sum_{x\in
L^\sharp}\Phi(\pi(x))\,p(x)\,q^{Q(x)}$$ is an element of $M_{r+\nu}(\rho)$.
\endproclaim

\subhead
Appendix
\endsubhead
We proof the theorem on representation by theta series.
It is consequence of the following
\proclaim{Lemma (Basic Transformation formula)}
Let $L$ be a lattice in a rational vector space $V$ of dimension $2r$,
let $Q$ be a
positive definite quadratic form on $V$ which takes on integral values on $L$,
let
$L^\sharp$ and $B$ be defined as above, let $w\in V\otimes\C$ with
$Q(w)=0$, let $\nu$ a non-negative integer, and let $z\in V$. Then one has
$$\align \tau^{-r-\nu}\sum\Sb x\in L\endSb &[B(w,x+z)]^\nu\,\e(-Q(x+z)/\tau)\\
&=\frac{i^{-r}}{\sqrt{[L^\sharp:L]}} \sum\Sb y\in L^\sharp\endSb
[B(w,y)]^\nu
\e(\tau Q(y)^t-B(y,z)),
\endalign$$
where $\tau$ is a variable in the complex upper half plane.
\endproclaim
The lemma is a well-known consequence of the Poisson summation
formula; for a proof cf\. \cite{\Schoen, p\. 206}. (For verifying that our
formula is
equivalent to the one given loc\. cit\. identify $L$ with $\Z^{2r}$
by choosing a $\Z$-basis $b_j$ of $L$, and note that then
$L^\sharp=G^{-1}\Z^{2r}$ and
$\det(G)=[L^\sharp:L]$ where $G=(B(b_j,b_k))$ is the Gram matrix of
$L$.
Moreover, the transformation formula loc\. cit\. is only stated for $\tau=it$
($t$ real); the general formula follows by analytic continuation.)

\demo{Proof of the theorem on representation by theta series}
Since any homogeneous spherical polynomial of degree $\nu$ can be written as
linear combination of the special ones $B(x,w)^\nu$ (where
$w\in\C$, $Q(w)=0$) we can assume that $p$ is of this special form.
Since $S$ and
$T$ generate $\SL$ it suffices to prove the asserted formula for these
elements.
For $A=T$ the formula is obvious. For proving the case $A=S$ let in the basic
transformation formula $z$ be an element of $L^\sharp$, multiply by $f(z)$ and
sum over a set of representatives $z$ for $L^\sharp/L$. Using $$\sum_{x\in
L^\sharp/L}\e(Q(x))=i^r\,\sqrt{[L^\sharp:L]}$$ (Milgram's theorem, e.g\.
\cite{\HuseMil, p\. 127}) we realize the claimed formula.  \qed\enddemo

\head
4. Theta series associated to quaternion algebras and the conformal characters
of the
five special model
\endhead
We now follow the procedure outlined in the foregoing section to construct
elements of $M_k(\rho_l)$ where $l$ denotes an odd prime
$l\equiv -1\bmod 3$ and $\rho_l$ is the matrix representation introduced in $\S
2$.

We first describe how to obtain $\rho_l$ from a proper Weil representation.
Let $\omega$ be the Weil representation associated to the quadratic module
$(\FF,\norm(x)/l)$. Here $\FF$
is the field with $l^2$ elements, and $\norm(x)=x\cdot\overline x$ with
$x\mapsto \overline x=x^l$  denoting the non-trivial automorphism of $\FF$.
Note that
$\tr(x\overline y)/l$ where
$\tr(x)=x+\overline x$ is the bilinear form associated to $n(x)/l$.
The Weil representation $\omega$ associated is thus a (right-)representation of
$\SL$ on the space
of functions $f\colon \FF@>>>\C$, and it is given by
$$f|\omega(T)(x)=\e(\norm(x)/l)\,f(x),\qquad
f|\omega(S)(x)=\frac{-1}l\sum\Sb y\in\FF\endSb\e(-\tr(\overline xy)/l)\,f(y).$$
Here we used
$$\sum_{x\in\FF}\e(\norm(x)/l)=-l,$$ as follows for instance from
Milgram's theorem and the considerations below where we shall obtain
$\FF=L^\sharp/L$
with a lattice of rank 4. Note that this identity implies in particular that
$\omega$ is a proper representation (cf\. the discussion in the preceding
section).

Let $\chi$ be one of the two characters of order 3
of the multiplicative group of nonzero elements in $\FF$, and let $G$ be the
subgroup of elements with $\norm(x)=1$.
Note that the existence of $\chi$ follows from the assumption $l\equiv -1\bmod
3$.
Let $X(\chi)$ be the subspace
of all $\phi\in X$ which satisfy $\phi(gx)=\chi(g)\phi(x)$ for all $g\in G$. It
is
easily checked that $X(\chi)$ is a $\SL$-submodule of $X$. In fact,
 it is even an irreducible one \cite{\NW, Satz 2}.
 As basis for $X(\chi)$ we may pick the functions $\chi_r$ ($1\le r\le l-1$)
which are defined by
$\chi_r(x)=\chi(x)$ if $\norm(x)=r$ and $\chi_r(x)=0$ otherwise. Let
$\Phi_\chi$ be the
complex column vector valued function on $\FF$ whose
$r$-th component equals $\chi_r$.
We then have $\Phi_\chi|A=\rho(A)\Phi_\chi$ with a unique
matrix representation $\rho\colon\SL\rightarrow
\GL(l-1,\C)$.
It is an easy exercise to verify the
identities
$$\rho(T)=\diag (\e^{2\pi i 1/l},\cdots,\e^{2\pi i (l-1)/l}),\quad
\rho(S)=\left(\lambda(rs)\right)_{1\le r,s\le l-1},$$
where we use
$$\lambda(r)=\frac{-1}{l}\sum\Sb x\in\FF\\ \norm(x)=r\endSb
\chi(x)\e(\tr(x)/l).$$
(In the identity $\norm(x)=r$ the $r$ has to be viewed as an element of $\FF$.)

Note that $\lambda(r)$ does not depend on the choice of $\chi$,
 as is easily deduced by replacing in its defining sum $x$ by
$\overline x$ and by using $\chi(\overline x)=\overline\chi(x)$ and
$\tr(\overline x)=\tr(x)$.
The independence of the choice of $\chi$ implies
that $\lambda(r)$, for any $r$, is contained in the field of $l$-th roots of
unities
(actually, $\lambda(r)$ is even real as follows from the easily proved facts
that $\rho(S)$
is unitary, symmetric and satisfies $\rho(S)^2=1$.)
Thus $\rho$ satisfies the properties listen in the Theorem characterizing
$\rho_l$ in \S\ 2, and hence is
equivalent to $\rho_l$. Indeed, by permuting the components of the vector
valued function $\xi_c$ occurring in
the definition of $\rho_l$ and by multiplying by a suitable diagonal matrix
we can even assume that $\rho=\rho_l$.

We now set $\Phi=\Phi_\chi+\Phi_{\overline\chi}$. The independence of
the matrices $\rho(A)$ ($A\in\SL$) of the choice of $\chi$ then implies
that the subspace spanned by the components of $\Phi$ is invariant under $\SL$,
and that
$\Phi|\omega(A)=\rho_l(A)\Phi$ for all $A\in\SL$. It is easily verified that
for all $x\in\FF$ one has
$$\Phi(x)\in\{0,-1,2\}^{l-1},\
(r\text{-th entry of }\Phi(x))\bmod l=\cases
\tr(x^{(l^2-1)/3})&\text{if }\norm(x)=r,\\
0&\text{otherwise},
\endcases
$$
where $r$ runs from 1 to $l-1$.

Next we describe lifts of $(\FF,\norm(x)/l)$.
Let $V$ be the quaternion algebra over $\Q$ ramified at $l$ and $\infty$. If we
set
$K=\Q(\sqrt{-l})$ then $V$ can be described as
$V=K+Ku$, where $u^2=-1/3$ and $\alpha u=u\overline\alpha$
for all $\alpha \in K$.
The map $c=\alpha+\beta u\mapsto \overline c:=\alpha-u\overline\beta$ defines
an
anti-involution of $V$. The reduced norm $\norm(c)$ and reduced trace
$\tr(c)$ of a $c\in V$ are given by
$$\norm(c)=c\overline
c=|\alpha|^2+\frac13|\beta|^2,\qquad \tr(c)=c+\overline
c=\alpha+\overline\alpha.$$

Let $\o$ be the ring of integers in $K$. Note that the rational
prime
$3$ splits in $K$ since $l\equiv -1\bmod 3$. i.e\.
$3=\p\overline\p$ with a prime ideal $\p$ in $K$. (Indeed, one can
take $\p=3\o+(1+\sqrt{-l})\o$.) We set
$$\O=\o+\p v,\qquad v=\cases
u&\text{for }l\equiv3\bmod4\\
\frac{1+u}2&\text{for }l\equiv1\bmod4
\endcases.
$$
It can be easily checked that $\O$ is an order in $V$ (i.e\. a subring
which, viewed as
$\Z$-module, is free of rank 4). In fact, $\O$ is even a maximal order since
the
determinant of the Gram matrix $(\tr(e_j\overline e_k))$, for any $\Z$-basis
$e_j$ of $\O$, equals
$l^2$ (cf\.  \cite{\Vi, Chap\. III, Corollaire 5.3}.).

We now have
\proclaim{Lemma}
(1) The dual lattice of $\sqrt{-l}\,\O$ w.r.t\. the quadratic form $\norm(c)/l$
is
$\O$. The quotient ring $\O/\sqrt{-l}\O$ is the field with $l^2$
elements, and the anti-involution $c\mapsto\overline c$ on $\O$
induces the Frobenius automorphism $x\mapsto x^l$ on $\O/\sqrt{-l}\O$.

(2) Let $I\subset\O$ be an $\O$-left ideal, and let $n=\norm(I)$ be the reduced
norm
of $I$ (i.e\. the g.c.d\. of the integers $\norm(x)$ where $x$ runs through
$I$).
Then the dual lattice of
$\sqrt{-l}I$ with respect to $\norm(c)/ln$ is $I$. There exists a $c_0\in I$
such that
$n(c_0)/n\equiv 1\bmod l$, and for any such $c_0$ the map $c\mapsto c_0\cdot c$
defines an isomorphism
of quadratic modules
$(\O/\sqrt{-l}\O,\norm(x)/l)@>\approx>>(I/\sqrt{-l}I,\norm(x)/ln)$.
\endproclaim

Here, for convenience, we use the same symbols $\overline n(x)/nl$ for the
quadratic form on $I$
 as well as for the quadratic form induced by it on $I/\sqrt{-l}I$.
The Lemma follows easily from standard facts in the theory of quaternion
algebras; for
the reader's convenience we sketch the proof in the Appendix to this section.

The Lemma provides us with lifts $(I,n(x)/nl)$ of $(\FF,\norm(x)/l)$, and we
now can write down
explicitly elements of $M_k(\rho_l)$.

To this end let
$\Phi\colon\FF=\O/\sqrt{-l}\O@>>>\{-1,0,2\}^{l-1}$ be defined as above.
Let $I$ be an $\O$-left ideal,
choose $c_0$ as in the Lemma, and let
$$\pi:I@>>>\O/\sqrt{-l}\O,\qquad\pi(c)=\lambda+\sqrt{-l}\O\quad
\text{with }c\equiv\lambda c_0\bmod\sqrt{-l}\O.$$

Finally, let $p$ be a  homogeneous spherical
polynomial function on $V$.
If we write polynomial functions on $V$ as polynomials $p$ in $\alpha$,
$\overline\alpha$, $\beta$, $\overline \beta$, then it is spherical of degree
$\nu$
(with respect to any nonzero multiple of $n(c)$) if and only if $p$ is
homogeneous
of degree $\nu$ and satisfies
$$
\big(\frac{\partial^2}{\partial\alpha\partial\overline\alpha}
+3\frac{\partial^2}{\partial\beta\partial\overline\beta}\big) p=0.
$$

Set
$$\theta(\tau;I,p)=\sum_{c\in I}\Phi(\pi(c))\,p(c)\,q^{n(c)/n(I)l}.$$
We suppress the dependence of this function on $c_0$ since a different choice
results only in
multiplying $\theta(\tau;I,p)$ by a scalar.
By the Theorem on Realization by theta series we then have
$$\theta(\tau;I,p)\in M_{2+\deg(p)}(\rho_l).$$

It is easy to compute these functions with aid of a computer. In fact, by a
computer
calculation we found

\proclaim{Theorem}
Let $l$, $k$ be as in Table 2 (in \S\ 2). Then the space $M_k(\rho_l)$
is spanned by the series $\theta(\tau;I,p)$, where $I=\O$ for $l\not=17$, and
$I=\O,\O\p$ for $l=17$,
and where $p$ runs through the homogeneous polynomial functions on the
quaternion algebra
$V$ of degree $k-2$ which are spherical with respect to the quadratic form
$\norm(c)$.
\endproclaim

It is an open question whether the spaces $M_k(\rho_l)$, for arbitrary $k$ or
primes $l$
($\equiv -1\bmod l$), are always spanned by theta series of the form
$\theta(\tau;I,p)$, or,
more generally, which spaces $M_k(\rho)$ of vector valued modular forms at all
can be generated by theta series.

As explained in section 2 we are especially interested in the one-dimensional
subspace
$M_k^{(\delta)}(\rho_l)$ of functions in $M_k(\rho_l)$ which
are $\BO(q^\delta)$ with $\delta$ as in Table 2. Here we have

\proclaim{Theorem (Theta formulas for conformal characters)}
(1) Let $c$, $l$, $k$ and $\delta$ be as in Table 2 (in \S\ 2), and let $I=\O$
for $l\not=17$
and $I=\O\p$ for $l=17$.
Then there exists a homogeneous spherical polynomial function $p$ of degree
$k-2$ such that the
$\theta(\tau;I,p)$ is nonzero and satisfies $\theta(\tau;I,p)=\BO(q^\delta)$.

(2) Moreover, for any $p$ with this property, there exists a nonzero constant
$\kappa$
 such that the components of
the Fourier coefficients of $\kappa\theta(\tau;I,p)$ are rational integers.
In particular, the components of $\kappa\eta(\tau)^{-2k}\theta(\tau;I,p)$
satisfy
the properties of conformal
characters (1) to (5) stated in \S\ 2.
\endproclaim

\demo{Proof}
(1) The existence of a $p$ with Fourier development starting at $q^\delta$
follows from
the preceding theorem and the fact that the subspace $M^{(\delta)}_k(\rho_l)$
contains
nonzero elements. For the latter cf\. the discussion in \S\ 2; of course, it
can
also be checked by a straight forward calculation using the $\theta(\tau;I,p)$
that
$M^{(\delta)}_k(\rho_l)$ is
one-dimensional.

(2) Because of the latter it is clear then that, for proving the second
statement of
the theorem, it suffices to prove that, for at least one $p$ satisfying the
condition
$\theta(\tau;I,p)=\BO(q^\delta)$, the function $\theta(\tau;I,p)$ has rational
Fourier
coefficients.

For proving this let $P_\nu(F)$ be the set of spherical homogeneous functions
$p$ on $V$
of degree $\nu$ which are defined over the subfield $F\subset\C$. By the latter
we mean
that the coefficients of $p(c)$, when written as polynomial in the coefficients
of $c$
with respect to a fixed basis of $V$, are in $F$. Note that this property does
not depend
on the choice of the $\Q$-basis of $V$. Since $P_\nu(F)$ is the kernel of a
differential
operator which has constant rational coefficients, when written with respect to
any
$\Q$-basis of $V$, it is clear that $P_\nu(\C)=P_\nu(\Q)\otimes\C$, i.e. we can
find a
basis of $P_\nu(\C)$ which is contained in $P_\nu(\Q)$. But then we deduce,
using the
preceding theorem, that $M_k(\rho_l)$ has a basis $\theta_j$ ($1\le j\le d$)
whose
Fourier coefficients $a_{\theta_j}(r)$ ($r=1,2,\dots$) are elements of
$\Q^{l-1}$.
For deducing this note that $\theta(\tau;I,p)$ for $p\in P_\nu(\Q)$ has
rational Fourier
coefficients since $\Phi(x)$ is rational. The elements of
$M^{(\delta)}_k(\rho_l)$ are
now the linear combinations $\sum_j c_j\theta_j$ such that $\sum_j c_j
a_{\theta_j}(r)=0$
for all $1\le r< l\delta$. Since the latter system of linear equations is
defined over
$\Q$ and has a nonzero solution by part (1) we conclude the existence of
rational nonzero
solution, i.e\. the existence of a linear combination of the $\theta_j$ with
Fourier coefficients in $\Q$.
\qed\enddemo

If we pick a $p$ as described in the theorem, and if we denote by $\xi_{c,h}$
the
$r$-th component of
$\eta^{-2k}\theta(\tau;I,p)$, where $\frac rl-\frac k{12}\equiv h-\frac
c{24}\bmod\Z$
then it is clear that these functions satisfy properties (1) to (5) of \S\ 2
(after multiplied by a constant, if necessary). Hence, by the uniqueness result
cited in section 1, they are up to a constant the conformal
characters of the $\W$-algebras introduced in the same section. In fact, the
$\xi_{c,h}$
($l\not=5$) have interesting product expansions, which we shall discuss
elsewhere;
from these product expansions it can immediately read off that
they can be normalized such that their Fourier coefficients are even
non-negative
integers, as its should be for conformal characters.

\subhead
Appendix
\endsubhead
\demo{Proof of the Lemma}
(1) Let $l\equiv-1\bmod4$. For $c=\alpha+u\beta\in V$ we have
$$\tr(c\overline\O\sqrt{-l})/l
=\tr(\alpha\,\o/\sqrt{-l})+\tr(\beta\,\overline\p/3\sqrt{-l}).$$
Thus the left hand side is in $\Z$
if and only if each of the two terms on the right are in
$\Z$. The latter is easily checked to be equivalent to
$\alpha\in\o$ and $\beta\in\p$, i.e\. to $c\in\O$. The case
$l\equiv1\bmod4$ can be treated similarly, and is left to the reader.

It is clear that
$\O/\sqrt{-l}\O$ is a ring of characteristic $l$ with $l^2$ elements.
Hence it is isomorphic to a ring extension of $\F(l)=\Z/l\Z$ with $l^2$
elements.
Moreover, it contains a root of $X^2+3$, namely $3u+\sqrt{-l}\O$. Since
$-3$ is not a quadratic residue modulo $l$ the polynomial $X^2+3$ is
irreducible over $\F(l)$, hence $\O/\sqrt{-l}\O$ is a field. The
anti-involution $c\mapsto\overline c$ induces an automorphism of the
field $\O/\sqrt{-l}\O$ which is nontrivial since it maps $u$ to $-u$, and
which hence is the Frobenius automorphism.

(2) If $I$ is an $\O$-left ideal then $I^*=\overline I\cdot\O^*/\norm(I)$,
where, for any left ideal $I$, we use
$$I^*=\{c\in V\,|\, \tr(Ic)\in\Z\}$$
(We were not able to find a reference for this basic formula: it can easily be
proved
using adelic methods. However, we shall need it only for $I=\O$ or $I=\O\p$
(cf\. the
two theorems of the preceding section), and here it can be easily verified by
direct
computation. We omit the details for the general case.) Thus we find
$\tr(\sqrt{-l}I\overline c)/\norm(I)l\in\Z$
if and only if $\overline c\in\overline I\cdot\O^*\sqrt{-l}$. Using
$\O^*=\O/\sqrt{-l}$,
as follows from part (1), we find that the latter statement is indeed
equivalent to $c\in I$.

Left-multiplication in the quaternion algebra induces on $I/\sqrt{-l}I$
a structure of a one-dimensional $\O/\sqrt{-l}\O$-vector space.
Let $c_0+\sqrt{-l}I$ be a basis element. Clearly $\norm(c_0)/\norm(I)$ is not
divisible by $l$
since otherwise $\norm(c)/\norm(I)$ would be divisible by $l$ for any $c\in I$
contradicting
the definition of $\norm(I)$ as g.c.d\. of all $\norm(c)$ ($c\in I$). Thus we
can choose
a $\lambda\in\O$ with $\norm(\lambda)\norm(c_0)/\norm(I)\bmod l$. Replacing
$c_0$
by $\lambda c_0$ it is then clear that $c\mapsto c c_0$ induces the claimed
isomorphism.
\qed\enddemo

\head
5. Comparison to formulas derivable from the representation theory of Kac-Moody
and Casimir
$\W$-algebra
\endhead

In this section we compare our explicit formulas for
the conformal characters with the ones
obtained from the representation theory of Casimir $\W$-algebras
\cite{\FreKaWa},
the Virasoro algebra \cite{\RoCa} and Kac-Moody algebras \cite{\Kac}.

The last three rational models in Table 2 are minimal models
of so-called Casimir-$\W$-algebras.
For this kind of algebras the minimal models have been determined
(assuming a certain conjecture) in \cite{\FreKaWa}.
The representation theory of the two composite rational models
($\W_{{\Cal G}_2}(2,1^{14})$ and $\W_{{\Cal F}_4}(2,1^{26})$)
is well-known \cite{\RoCa,\Kac}.

In order to give the explicit formulas for the conformal characters of the
minimal models of the Casimir-$\W$-algebras, the Virasoro algebra and
Kac-Moody algebras we have to fix some notation first.

Let $\Cal K$ be a simple complex Lie algebra of rank $l$ and dimension $n$,
$h \ ({\check h})$ its (dual) Coxeter number,
$\rho \ ({\check \rho})$ the sum of its (dual) fundamental
weights, $W$ the Weyl group and $\Lambda$ (${\check \Lambda}$)
the (dual) weight lattice of  $\Cal K$. For $\lambda \in \Lambda$
let $\pi_{\lambda}$ denote the highest weight representation
with highest weight $\lambda$.

Firstly, consider the case of the three rational models of the
Casimir-$\W$-algebras.
Formulas for the central charge, the conformal dimensions and the conformal
characters of rational models of Casimir $\W$-algebras have been derived
assuming a
certain conjecture \cite{\FreKaWa, p.\ 320}:
$$\align
   c &= c(p,q) = l - {12\over pq}(q\rho - p{\check \rho})^2, \\
   h_{\lambda,{\check \nu}} &= {1\over 2pq}
     \bigl( (q(\rho+\lambda)-p({\check \rho}+{\check \nu}))^2-
            (q\rho-p{\check \rho})^2 \bigr),\\
  \chi_{\lambda,{\check \nu}}(q)  &=
      \eta(q)^{-l} \sum_{w \in W}
      \sum_{ t \in {\check \Lambda}} \epsilon (w)
      q^{{1\over2pq}{(q w(\lambda+\rho) - p ({\check \nu}+{\check \rho}) +
pqt)}^2}
\endalign
$$
where $p,q$ are coprime integers satisfying
${\check h} \le p, h \le q$ and
where $\lambda \in \Lambda$ and ${\check \nu} \in {\check \Lambda}$
are so-called dominant integral weights (the abstract form of the finitely many
$\lambda$,
${\check \nu}$ is described in  \cite{\FreKaWa}; their explicit form can e.g.\
be found in Appendix
D of \cite{\BEHHH}).

Using these formulas for ${\Cal B}_2$ with $c(p,q) = c(11,6) = -{444\over11}$
and for ${\Cal G}_2$ with $c(p,q) = c(17,12) = -{1420\over17}$ for $\W(2,4)$
and $\W(2,6)$, respectively, one obtains the conformal characters given in
the last section (as can be checked by simply comparing enough Fourier
coefficients).
The last rational model, of type $\W(2,8)$, is a rational model of
${\Cal WE}_7$ with $c(p,q) = c(18,23)$. However, in this case
the above formula for the corresponding conformal characters contains a
sum over a rank 7 lattice (the dual weight lattice) and a
sum over the Weyl group of ${\Cal E}_7$ which has order $2.903.040$.
Therefore, this formula is of no practical use for explicit calculations in
this case. However, our formula in the foregoing section involves only a sum
over a rank 4 lattice which is easy to implement on a computer.

Secondly, consider the rational models
$\W_{{\Cal G}_2}(2,1^{14})$ and $\W_{{\Cal F}_4}(2,1^{26})$.
These rational models are ``tensor products'' of the Virasoro minimal
model with $c=-{22\over5}$ and the rational model associated to the level 1
Kac-Moody algebra  of ${\Cal G}_2$ or ${\Cal F}_4$, respectively.
The two conformal characters of the Virasoro minimal model with central charge
$c=-{22\over5}$  are given by \cite{\RoCa}
$$  \chi_{0}^{Vir}(q) =
    q^{\frac{11}{60}}\prod_{n \equiv \pm 2 \bmod 5}(1-q^n)^{-1},\qquad
    \chi_{-1/5}^{Vir}(q) =
    q^{-\frac1{60}}  \prod_{n \equiv \pm 1 \bmod 5}(1-q^n)^{-1}.
$$

The characters of rational models associated to the level 1 Kac-Moody
algebras are well known from the Kac-Weyl formula \cite{\Kac, p\. 173}.
The rational model associated to the level
$k$ Kac-Moody algebra of $\Cal K$ has the following central charge
and conformal dimensions
$$ c^{\Cal K}(k) = {12k\over {\check h}({\check h}+k)}\rho^2,
   \qquad h^{\Cal K}_{\lambda} =  {(\rho+\lambda)^2-\rho^2 \over
                             2({\check h}+k) }
   \quad( (\lambda, \psi) \le k)
   $$
where $\psi$ is the highest root of $\Cal K$.
The corresponding characters read
$$ \chi^{{\Cal K},\lambda}(q) =
   \eta(q)^{-n} q^{{n-c^{\Cal K}(k)\over24}}\sum_{t \in {\check \Lambda}}
                     \dim(\pi_{\rho+\lambda+({\check h}+k)t})
                     q^{ (\rho+\lambda+({\check h}+k)t)^2-\rho^2
                         \over 2({\check h}+k) }.
$$
The two conformal characters associated to the level 1 Kac-Moody algebras
of ${\Cal G}_2$ and ${\Cal F}_4$ are given by:
$$ \chi_{0}^{\Cal K} = \chi^{{\Cal K},0} \quad
   \chi_{h}^{\Cal K} = \chi^{{\Cal K},\lambda_1}
$$
where $h=h^{\Cal K}_{\lambda_1}=2/5$ or $3/5$ and  $\lambda_1$ is
the fundamental weight of ${\Cal G}_2$ or ${\Cal F}_4$
with $\dim(\pi_{\lambda})=7$ or $26$, respectively.

Using these formulas one obtains exactly the four conformal characters of the
models $\W_{{\Cal G}_2}(2,1^{14})$ and $\W_{{\Cal F}_4}(2,1^{26})$:
$$\chi_{0}     = \chi_{0}^{Vir}     \cdot \chi_{0}^{\Cal K}, \quad
  \chi_{-1/5}  = \chi_{-1/5}^{Vir}  \cdot \chi_{0}^{\Cal K}, \quad
  \chi_{h}     = \chi_{0}^{Vir}     \cdot \chi_{h}^{\Cal K}, \quad
  \chi_{h-1/5} = \chi_{-1/5}^{Vir}  \cdot \chi_{h}^{\Cal K}
$$
with ${\Cal K} = {\Cal G}_2,{\Cal F}_4 $ and $h=2/5,3/5$ respectively.
The product formulas for the Virasoro characters and the formula for
the conformal characters associated to the Kac-Moody algebras show that
the Fourier coefficients of the two rational models are positive integers.
Indeed, as one can show by comparing enough Fourier coefficients,
these conformal characters are equal to the ones computed in the last
section.

\head
Acknowledgments
\endhead

W. E. would like to thank the research group of W. Nahm for
many useful discussions. Parts of the present article were written by the
second
author during his membership at the Mathematical Sciences Research Institute at
Berkeley, and he would like to thank the staff of the MSRI for providing such a
warm and fruitful atmosphere. All computer calculations have been performed
with
the computer algebra package PARI-GP ~\cite{\GP}.


\enddocument